\documentclass[12pt,preprint]{aastex}

\shorttitle{Dust Extinction on GRB Afterglows}
\shortauthors{L\"{u} et al.}

\begin{document}

\title{Effect of Dust Extinction on Gamma-ray Burst Afterglows}

\author{\sc Gu-Jing L\"{u}\altaffilmark{1}, Lang Shao\altaffilmark{1,2}, Zhi-Ping Jin\altaffilmark{1}, and Da-Ming Wei\altaffilmark{1}}
\altaffiltext{1}{Purple Mountain Observatory, Chinese Academy of Sciences, Nanjing 210008, China}
\altaffiltext{2}{Department of Physics, Hebei Normal University, Shijiazhuang 050016, China}
\email{lang@pmo.ac.cn(L.S.)}

\begin{abstract}

In order to study the effect of dust extinction on the afterglow of gamma-ray
bursts (GRBs), we carry out numerical calculations with high precision based on rigorous
Mie theory and latest optical properties of interstellar dust grains, and analyze the different
extinction curves produced by dust grains with different physical parameters. Our results
indicate that the absolute extinction quantity is substantially determined by the medium
density and metallicity. However, the shape of the extinction curve is mainly determined by
the size distribution of the dust grains. If the dust grains aggregate to form larger ones, they
will cause a flatter or grayer extinction curve with lower extinction quantity. On the contrary,
if the dust grains are disassociated to smaller ones due to some uncertain processes, they
will cause a steeper extinction curve with larger amount of extinction. These results might
provide an important insight into understanding the origin of the optically dark GRBs.

\end{abstract}


\section{INTRODUCTION} \label{sec:intro}

Gamma-ray burst (GRB) is known as one of the most energetic
stellar explosions in the universe. At present, the {\it Swift}
satellite \citep{gehrels04}, a NASA mission dedicated to monitor this phenomenon,
carries three instruments with separate wave bands: Burst Alert
Telescope (BAT; $\sim15-150$~keV), X-Ray Telescope (XRT; $\sim
0.3-10$~keV) and Ultraviolet/Optical Telescope (UVOT). BAT is
able to catch about 100 GRBs per year, and XRT is able to follow
them rapidly in the X-ray band and pinpoint their positions
accurately. However, UVOT could only detect the optical afterglows
in about $60\%$ of them. Some of the detected optical afterglows are
also weaker than those predicted by theoretical models. Those that
have weaker or no optical afterglow are called optically dark
GRBs \citep{vanderHorst09}.

The origin of dark GRBs is an open question in the research field of
GRBs. Generally, the extinction by ambient dust grains is considered
as the answer \citep{stratta04,schady07}, but the detailed theoretical implications are
uncertain. The extinction curves measured for GRB afterglows are
very diverse. Some of them are similar to that of the Small
Magellanic Cloud (SMC), i.e., a steep extinction curve; some of them
are similar to that of the Milky Way (MW), i.e., an extinction curve
with significant extinction bump at $\sim 2175~{\rm \AA}$ \citep{stratta04};
some of them are similar to that of a normal active galaxy with a
flat extinction curve, i.e., gray extinction \citep{stratta05,chen06}. Obviously, the
various extinction properties can not be fitted with the modeling
sample of existing extinction curves. It is still a mystery how to
infer the physical implications with different extinction curves. In
this work, we adopt the latest algorithm for calculating dust
physics and focus on the effect of dust extinction on GRB
afterglows. By analyzing the influences of different physical
parameters of dust grains on the extinction curves, we can reproduce
various observed extinction curves and provide theoretical basis for
the studies of dark GRBs and gray extinction.

\section{PHYSICAL MECHANISM OF DUST EXTINCTION} \label{sec:mech}
Considering a spherical dust grain with radius $a$ and complex
refractive index $\tilde{m}$, and based on Mie theory \citep{vandeHulst57}, the
extinction cross-section for the incident light with wavelength
$\lambda$ is
\begin{equation}\label{extcs}
    \sigma_{\rm ext}={2\pi\over k^2}\Sigma_{n=1}^\infty (2n+1){\rm Re}\{a_n+b_n\}\,,
\end{equation}
the scattering cross-section is
\begin{equation}\label{scacs}
    \sigma_{\rm sca}={2\pi\over k^2}\Sigma_{n=1}^\infty (2n+1)(|a_n|^2+|b_n|^2)\,,
\end{equation}
and the absorption cross-section is $\sigma_{\rm abs}=\sigma_{\rm
ext}-\sigma_{\rm sca}$, where $k=2\pi/\lambda$, and the scattering
coefficients $a_n$ and $b_n$ are
\begin{equation}\label{an}
    a_n={\tilde{m}\psi_n(\tilde{m}x)\psi_n'(x)-\psi_n(x)\psi_n'(\tilde{m}x) \over \tilde{m}\psi_n(\tilde{m}x)
    \xi_n'(x)-\xi_n(x)\psi_n'(\tilde{m}x)}\,,
\end{equation}
\begin{equation}\label{bn}
    b_n={\psi_n(\tilde{m}x)\psi_n'(x)-\tilde{m}\psi_n(x)\psi_n'(\tilde{m}x)\over \psi_n(\tilde{m}x)\xi_n'(x)-
    \tilde{m}\xi_n(x)\psi_n'(\tilde{m}x)}\,,
\end{equation}
where $x=ka=2\pi a/\lambda$ is the dimensionless size parameter, and
$\psi_n(x)$ and $\xi_n(x)$ are Riccati-Bessel functions.

Methods for numerical calculations based on Mie theory are mature
now. Owing to the rapid development of computer science, the
calculation of the infinite sums can be performed on popular PCs at
present, instead of using supercomputers in the past. Numerical
analyses indicate that the infinite series summation in
Equations~(1)-(4) can be approximated by the first $N=x+4x^{1/3}+2$
terms with a sufficiently high precision \citep{wiscombe80}, where $x$ is the
dimensionless size parameter mentioned above. Currently, there are a
few popular FORTRAN codes \citep{wiscombe80,bohren83}, which can be very efficient (for
a single calculation with $x\approx10^3$, it only takes a couple of
seconds on an Intel PC with a main frequency of 2.6 GHz), but is
also numerically unstable and can be very time-consuming for
multi-wavelength calculations, due to the lack of real-time
adjustment of the precision, especially when $x$ is very large (for
X-rays and large dust grains, $x\gtrsim 10^5$). For the evaluation
of absorption and scattering cross-sections over a broad bandpass
and a wide size range of dust grains, multiple analytical
approximations are usually adopted for interpolation \citep{wiscombe80}. In order
to study the properties of X-ray scattering and absorption by large
dust grains, we make an extensive use of the latest MieSold code in
the advanced language Mathematica which can make real-time
adjustment of the precision with a sacrifice of the speed (for a
single calculation with $x\approx10^4$, it takes about $1\,{\rm h}$
on an Intel PC with a main frequency of 2.6 GHz). Nevertheless, the
precision is greatly improved by self-adapting calculations over a
much larger parameter space\footnote{Zimmer C, Aragon S R,
Mie Scattering and Absorption from Bubbles and Spheres, Mathematica
Journal, to be submitted.}.

There are a variety of substances in the interstellar medium (ISM).
The composition and optical properties of most dust grains can not
be obtained directly by experiment or observation. They are mainly
measured jointly by laboratory experiments, theoretical modelings
and astronomical observations. At present, silicate and graphite are
known as the two most important ingredients in ISM \citep{draine03b}. Their
optical properties have been systematically studied by Draine and
his colleagues and the latest results on their complex refractive
indices $\tilde{m}$ have been summarized in Figure~1 \citep{draine84,laor93,li01,draine03a}. The
optical properties of graphite are highly anisotropic, and the value
of $\tilde{m}$ is dependent on the angle included between the
direction of electric field and the crystal axis. The ``$1/3-2/3$''
approximation is usually adopted in most evaluations, i.e., graphite
is assumed as a mixture of two types of isotropic substances. Among
them, $1/3$ constituent resembles the graphite with the electric
field parallel to the crystal axis and $2/3$ constituent resembles
the graphite with the electric field perpendicular to the crystal
axis \citep{draine93}. The X-ray edge absorption is taken into account in the
results shown in Figure~1 (see ${\rm Im}(\tilde{m})$, the imaginary
part of $\tilde{m}$). The edge absorptions of silicate are quite
abundant, including the multiple edge absorptions from Mg, Fe, Si
and O, while graphite has only a K edge absorption between 282 and
310 eV \citep{draine03a}.

Based on the above-mentioned optical properties of silicate and
graphite, we can obtain their absorption and scattering
cross-sections as functions of the grain size $a$ and the energy of
the incident light $E$ by precise evaluations according to Mie
theory. As shown in Figure~2, the results given by the MieSolid code
(Rigorous Mie) are very consistent with those previously combined
results based on multiple analytical approximations (Mie,
Rayleigh-Gans [RG] and Geometric Optics [GO]) and have higher
spectral resolution with better performance on the edge absorptions.
The only flaw is that it is unstable in the ultraviolet and soft
X-ray band when the grain size is larger than $1\,\mu{\rm m}$, which
shall be tackled in the future code debugging. In this work, we
adopt the existing approximative results for the unstable region via
interpolation which will have negligible effect on the final
results, since large grains in most standard dust models are
deficient.

Our results (as in Figure~2) indicate that, when the typical grain
size is small ($a<0.1\,{\rm \mu m}$) absorption dominates the
extinction with most of the incident energy transformed into ambient
thermal energy. In this case, the scattering will be relatively
weak, and it has been overlooked in previous works. On the contrary,
when the typical grain size is large ($a>0.1\,{\rm \mu m}$), the
scattering dominates the extinction, especially the X-ray scattering
will play an important role. As the grain is getting larger, the
scattering is more effective. These results might have crucial
implications for the studies of X-ray scattering in GRB
afterglows \citep{shao07,shao08}.
    
\begin{figure}[tbph]
\begin{center}
\includegraphics[scale=0.7]{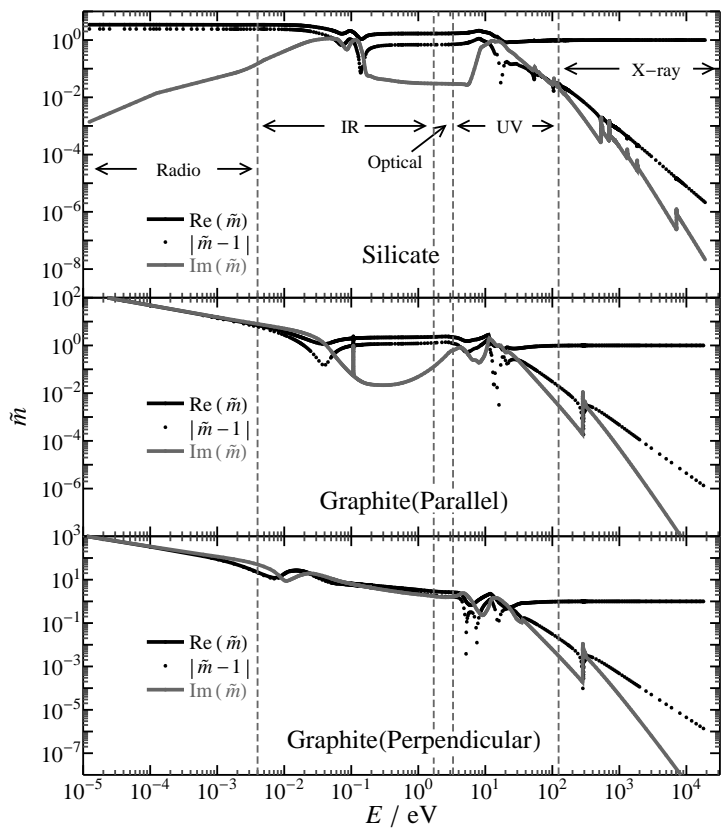}
\end{center}
\caption{Complex refractive indices $\tilde{m}$ of spherical silicate and graphite as
      functions of the energy of an incident photon. The real part and imaginary parts of the complex refractive indices
      represented by black and gray solid lines, respectively, and the values of $|\tilde{m}-1|$ that is frequently used in literatures
      are represented by dotted lines. }
\end{figure}

\section{EXTINCTION OF GRB AFTERGLOWS} \label{sec:ext}

The size distribution of the dust grains around GRBs \citep{mathis77} can
be assumed obey a power law between $(a_{\rm min},a_{\rm max})$
given by
\begin{equation}\label{size}
    {{\rm d}N_i\over {\rm d}a}(a)=A_i\times N_{\rm H}a^{\beta} \quad (a_{\rm min}\leq a\leq a_{\rm max})\,,
\end{equation}
where ${\rm d}N_i/{\rm d}a$ are the column densities per unit radius
of silicate $(i=1)$ and graphite $(i=2)$, respectively, $A_1$ and
$A_2$ are the coefficients that quantify their absolute column
densities, $N_{\rm H}$ is the column density of hydrogen atoms,
$\beta$ is the dimensionless power-law index. Accordingly, the dust
grain mass per unit hydrogen mass \citep{laor93}, i.e., the equivalent
metallicity is given by
    
\begin{figure}[tbph]
\begin{center}
\includegraphics[width=2.7in]{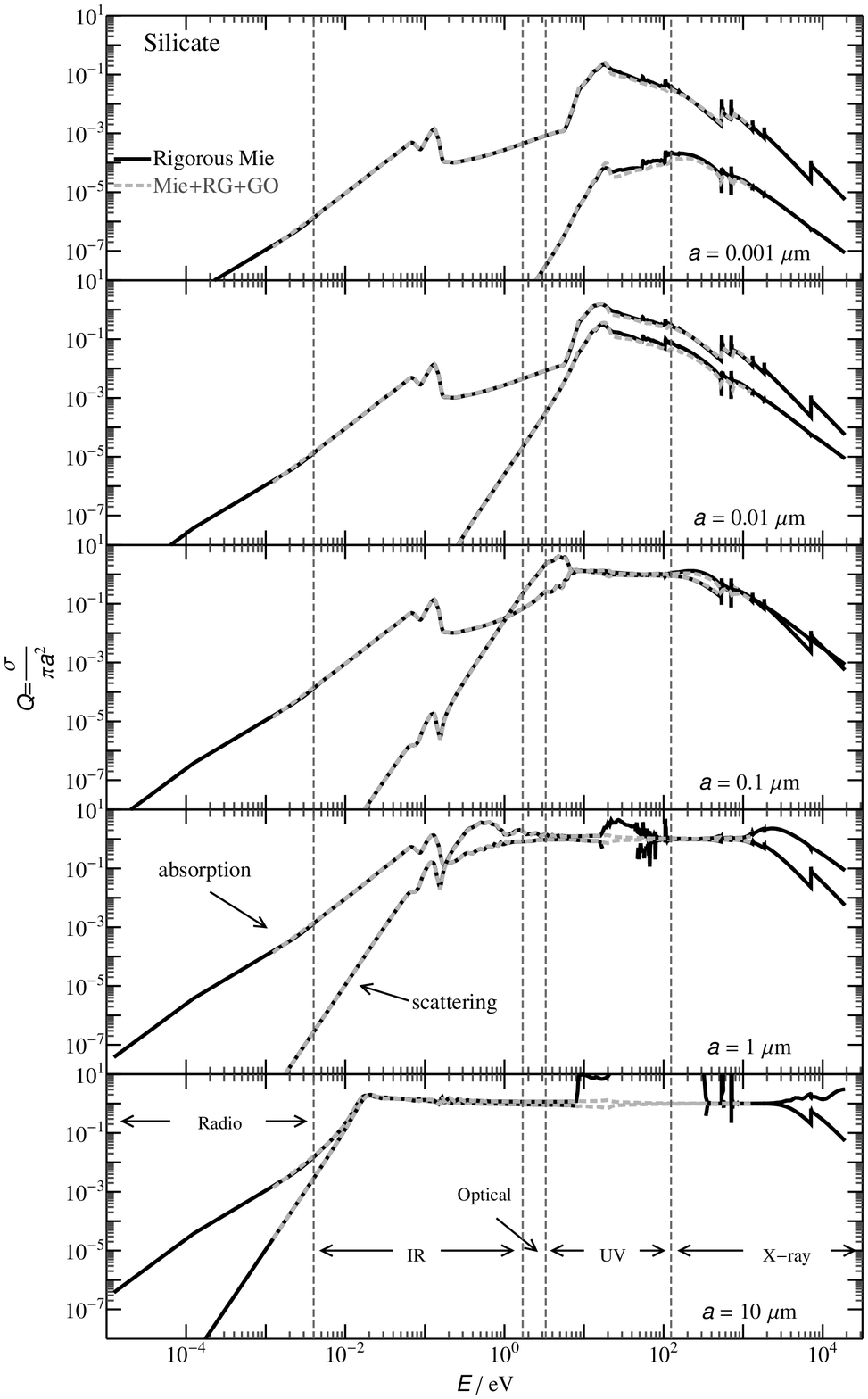}\includegraphics[width=2.7in]{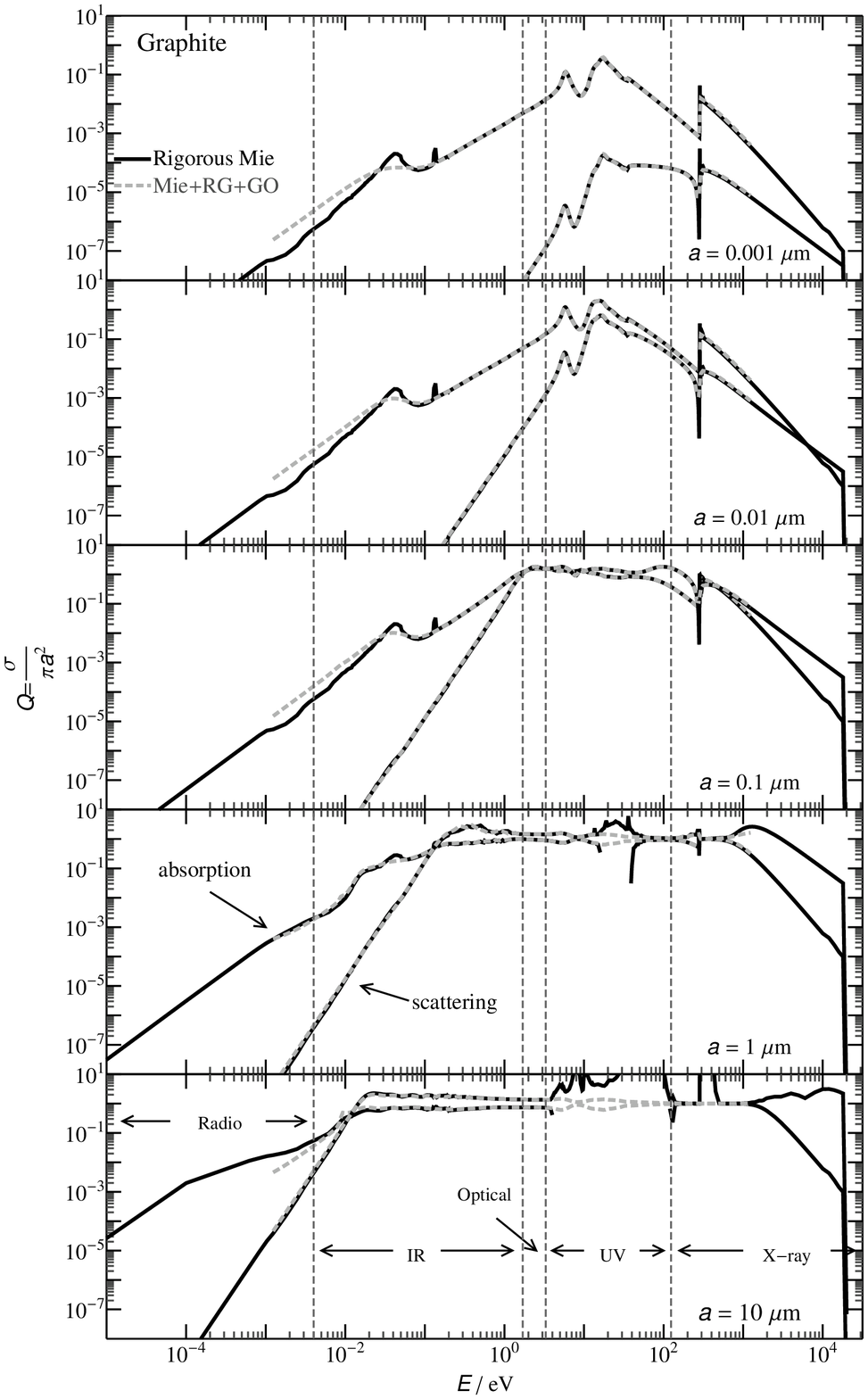}
\end{center}
\caption{Variations of absorption and scattering cross sections of spherical silicate and graphite with frequency. }
\end{figure}

\begin{equation}\label{fd}
    f_{\rm d}={4\pi\over 3 m_{\rm H}}{a_{\rm max}^{\beta+4}\over \beta+4}
    \left[1-\left({a_{\rm min}\over a_{\rm max}}\right)^{\beta+4}\right]\sum_i A_i \rho_i\,,
\end{equation}
where $\rho_1=3.3\,{\rm g/cm^3}$ and $\rho_2=2.3\,{\rm g/cm^3}$ are
the mass densities of silicate and graphite, respectively, $m_{\rm
H}$ is the mass of hydrogen atom. Here $f_{\rm d}$, $A_1$ and $A_2$
are not completely independent. Thereafter $f_{\rm d}$ and $A_1/A_2$
will be considered as two independent parameters. Meanwhile, the
extinction optical depth can be given by
\begin{equation}\label{A}
    \tau(\lambda)=\int  \sum_i \sigma_{\rm ext}^i(\lambda){{\rm d}N_i\over {\rm d}a}{\rm d}a\,,
\end{equation}
and the extinction magnitude is $A(\lambda)=1.086 \tau(\lambda)$.

The GRB afterglows that are emitted by the shock-accelerated
electrons in the relativistic outflow usually exhibit a power-law
spectrum from the optical to X-ray band, which hereafter is assumed
to be $F_\nu\propto\nu^{-1}$ (as shown by the gray solid line in
Figure~3; \citet{shao10}). The column density of hydrogen atoms is the
principal quantity that dominates the extinction from ultraviolet to
soft X-ray band. As shown in Figure~3, from top to bottom, the
solid, dotted, short-dashed and long-dashed lines represent that the
values of $N_{\rm H}$ are $10^{20}\,{\rm cm^{-2}}$, $10^{21}\,{\rm
cm^{-2}}$ , $10^{21.6}\,{\rm cm^{-2}}$ and $10^{22}\,{\rm cm^{-2}}$,
respectively. All the other physical parameters have the typical
vales in ISM, where the metallicity $f_{\rm d}$ is 0.01, the ratio
of silicate and graphite $A_1/A_2$ is 1 and the parameters for grain
size distribution are $\beta=-3.5$, $a_{\rm min}=0.005\,{\rm \mu m}$
and $a_{\rm max}=0.25\,{\rm \mu m}$. In general, the column density
of hydrogen atoms $N_{\rm H}$ determines the absolute amount of
extinction. In some dense regions of the surrounding medium $N_{\rm
H}$ could be very high (usually $N_{\rm H}>10^{22}\,{\rm cm^{-2}}$),
the optical to soft X-ray emissions from GRB afterglows would be
severely attenuated.

\begin{figure}[tbph]
\begin{center}
\includegraphics[width=4in]{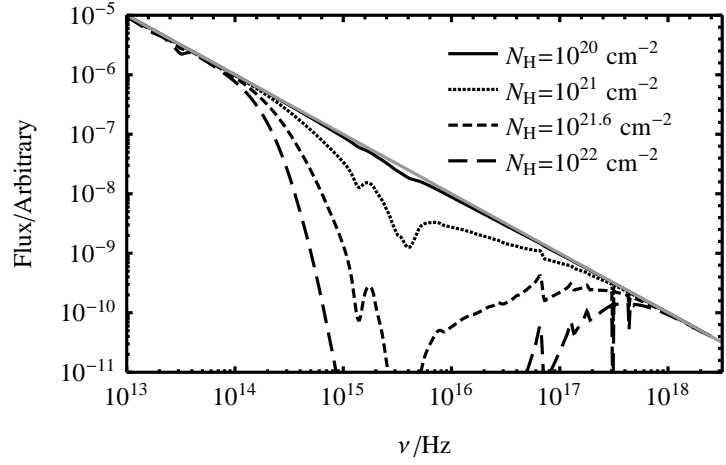}
\end{center}
\caption{Extinction of GRB afterglow by circum-stellar dust grains with different column densities of hydrogen
      nuclei $N_{\rm H}$. Gray straight line represents the intrinsic spectrum of the
      afterglow. }
\end{figure}

Besides the column density of hydrogen atoms, many other factors
will also affect the extinction curve (including the absolute amount
of extinction and the profile of the extinction curve). Herein we
mainly consider some key physical quantities: the ratio of silicate
and graphite $A_1/A_2$, the metallicity $f_{\rm d}$, parameters for
dust grain size distribution $\beta$ and $a_{\rm max}$. The impact
of $A_1/A_2$ on the extinction curve is shown in Figure~4, where
$A_1/A_2=0.6$, $1.0$, $1.6$ and $2.0$ are represented by the solid,
dotted, short-dashed and long-dashed lines, respectively. The other
parameters also have the typical values, i.e., $f_{\rm d}=0.01$,
$N_{\rm H}=10^{21.3}\,{\rm cm^{-2}}$, $\beta=-3.5$, $a_{\rm
min}=0.005\,{\rm \mu m}$ and $a_{\rm max}=0.25\,{\rm \mu m}$. As
revealed in the figure, $A_1/A_2$ mainly affects the extinction bump
around $2175\,{\rm \AA}$, which has been known to be caused by small
graphite grains \citep{draine93}. Therefore, as $A_1/A_2$ increases, the
extinction bump gets flatter. In general, the composition of dust
grains has weak effect on the extinction curve and can not account
for why we can observe evidently different extinction curves from
GRB afterglows.

\begin{figure}[tbph]
\begin{center}
\includegraphics[bb=17 0 224 142, width=2.5in]{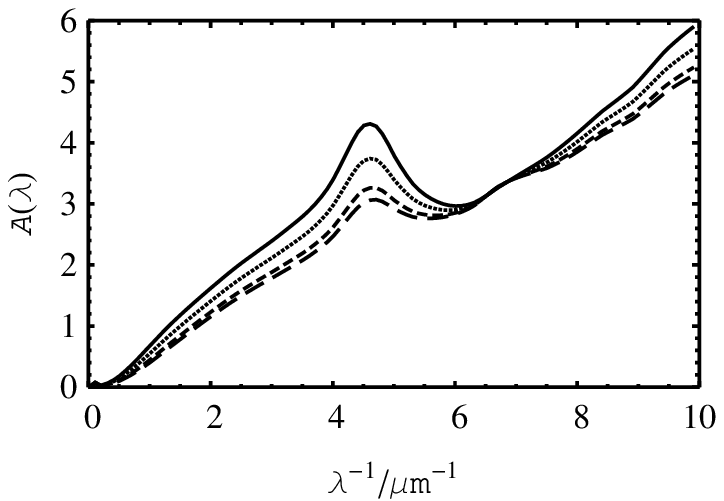}\includegraphics[width=2.5in]{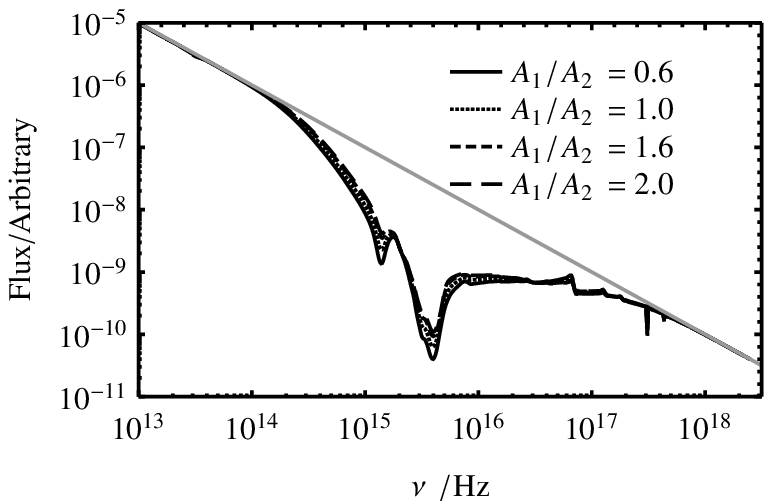}
\end{center}
\caption{Extinction of GRB afterglow by circum-stellar dust grains
   with different relative abundances between silicate and graphite $A_1/A_2$. Left panel is the
    extinction curve, and right panel is the attenuated afterglow spectrum. Gray straight
      line in the right panel represents the intrinsic spectrum of the afterglow. }
\end{figure}

The metallicity $f_{\rm d}$ has a great effect on the extinction
curve. Being similar to the column density of hydrogen atom $N_{\rm
H}$, which dominates the absolute amount of extinction, larger
metallicity causes stronger extinction. As shown in Figure~5,
$f_{\rm d}=0.001$, $0.004$, $0.007$ and $0.01$ are represented by
the solid, dotted, short-dashed and long-dashed lines, respectively.
The other parameters also have their respective typical values,
i.e., $A_1/A_2=1$, $N_{\rm H}=10^{21.3}\,{\rm cm^{-2}}$,
$\beta=-3.5$, $a_{\rm min}=0.005\,{\rm \mu m}$ and $a_{\rm
max}=0.25\,{\rm \mu m}$. The ambient environment around a GRB is
very complicated. There might be a high metallicity if the explosion
occurs in the latter phase of the massive progenitor star which is
an ideal place for the dust formation. This might be the leading
cause of the severe extinction and the optically dark GRBs.

Obviously, as the computing results indicate, the power-law index of
the dust grain size distribution $\beta$ mainly determines the
profile of the extinction curve. The chief reason is that dust
grains with different sizes have different contributions to the
extinction at different photon frequencies. This is governed by the
physics of dust scattering, which is weakly affected by the
ingredients of dust grains. As shown in Figure~6, $\beta=-3.5$,
$-2.5$, $-1.5$ and $-0.5$ are represented by the solid, dotted,
short-dashed and long-dashed lines, respectively. The other
parameters also have the typical values, i.e., $f_{\rm d}=0.01$,
$A_1/A_2=1$, $N_{\rm H}=10^{21.3}\,{\rm cm^{-2}}$, $a_{\rm
min}=0.005\,{\rm \mu m}$ and $a_{\rm max}=0.25\,{\rm \mu m}$. As the
size of dust grains increases, the number of small dust grains
decreases and the extinction bump around $2175\,{\rm \AA}$ also
becomes less evident. Another interesting feature is that $\beta$
barely affects the optical extinction $A_{\rm V}$. This may explain
why we can usually observe different extinction curves from GRB
afterglows, but $A_{\rm V}$ is barely correlated with $N_{\rm
H}$ \citep{schady07}. Our computing results indicate that difference of size
distribution of dust grains might be the internal cause.
    
\begin{figure}[tbph]
\begin{center}
\includegraphics[bb=18 0 223 143, width=2.5in]{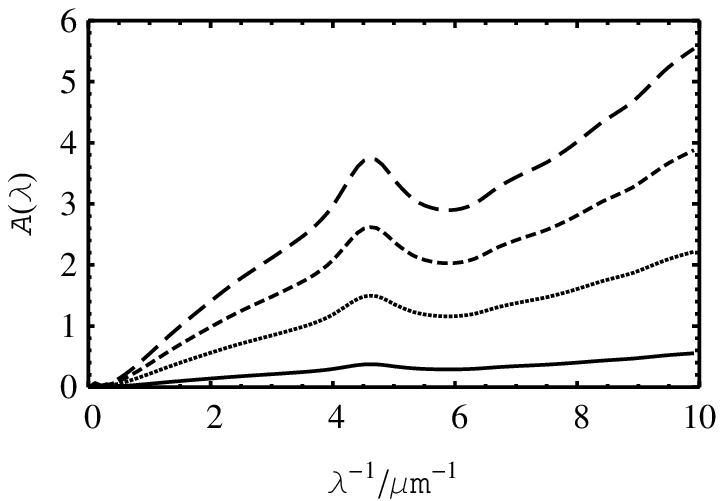}\includegraphics[width=2.5in]{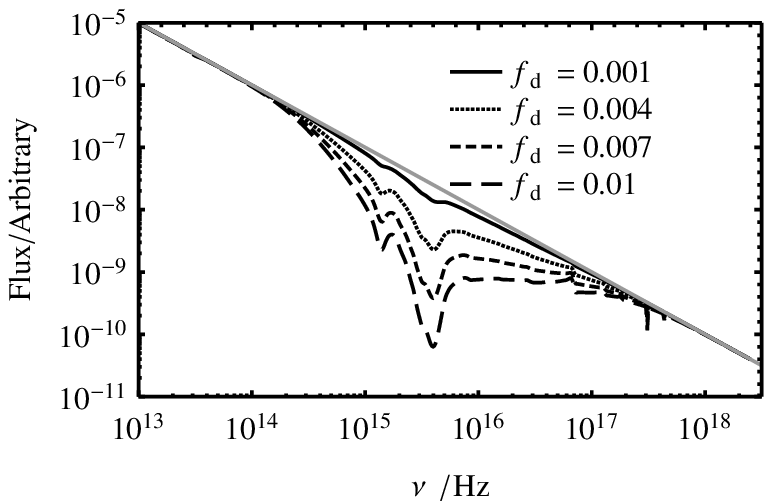}
\end{center}
\caption{Extinction of GRB afterglow by circum-stellar dust grains
   with different metallicities. Left panel is the
      extinction curve, and right panel is the attenuated afterglow spectrum. Gray
      straight line in the right panel represents the intrinsic spectrum of the afterglow. }
\end{figure}

The upper limit for the size distribution of dust grains $a_{\rm
max}$ is also an important parameter that determines the absolute
amount of extinction. Being different from the above-mentioned
column density of hydrogen atom $N_{\rm H}$ and the metallicity
$f_{\rm d}$, as $a_{\rm max}$ increases, the absolute amount of
extinction decreases. As shown in Figure~7, the relations $a_{\rm
max}=10^{-0.5}$, $1$, $10^{0.5}$ and $10\,{\rm \mu m}$ are
represented by the solid, dotted, short-dashed and long-dashed
lines, respectively. The other parameters still have the typical
values, i.e.,

\begin{figure}[tbph]
\begin{center}
\includegraphics[bb=18 0 223 143, width=2.5in]{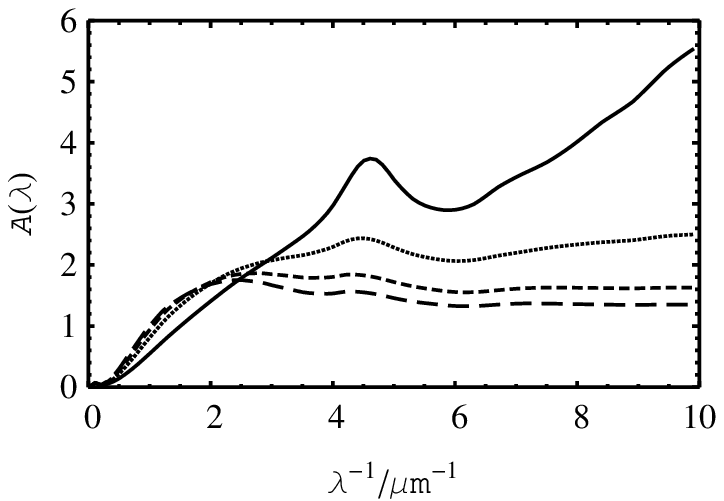}\includegraphics[width=2.5in]{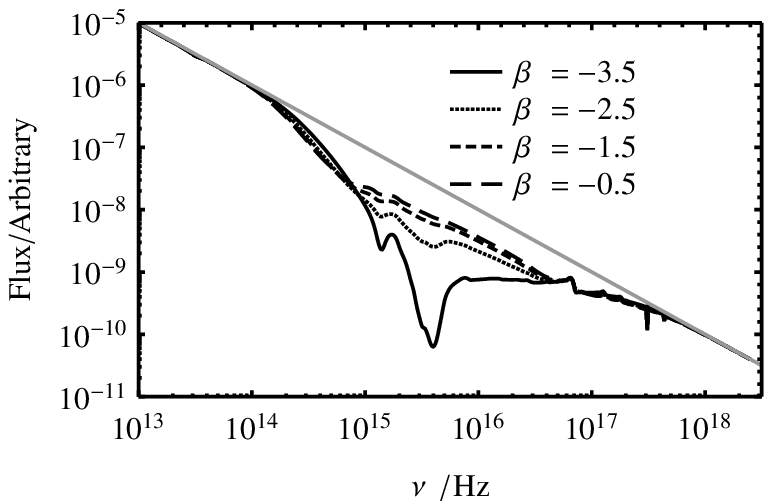}
\end{center}
\caption{Extinction of GRB afterglow by circum-stellar dust
   grains with different indices of size distribution $\beta$. Left panel is the
      extinction curve, and right panel is the attenuated afterglow spectrum. Gray
      straight line in the right panel represents the intrinsic spectrum of the afterglow. }
\end{figure}
    
 $f_{\rm d}=0.01$, $A_1/A_2=1$, $N_{\rm
H}=10^{21.3}\,{\rm cm^{-2}}$, $\beta=-3.5$ and $a_{\rm
min}=0.005\,{\rm \mu m}$. This phenomenon is due to an underlying
assumption in our calculations that the total mass of the dust
grains is conserved. Larger dust grains are formed by the
aggregation of smaller ones. As the number of larger dust grains
increase, the total number density of the dust grains will naturally
decrease. Our computing results indicate that the absolute amount of
extinction will remarkably decrease, and as the size of dust grain
increases, the extinction curve will become flatter, causing gray
extinction \citep{stratta04,li08}. Therefore, with $\beta$ and $a_{\rm max}$ both
varying, we would expect that the dust grains with typically larger
sizes will cause weak extinction and have a flatter extinction
curve, i.e., causing gray extinction. This explains why most
observed optically bright afterglows exhibit flat extinction curves \citep{stratta04}.
On the contrary, the dust grains with typically smaller
sizes would cause more severe extinction, i.e., causing optically
dark bursts, and have remarkably steeper extinction curves.
Therefore, numerous computing results indicate that the discrepancy
and evolution of the sizes of dust grains can have very crucial
effects on the extinction curves of GRB afterglows.
    
\begin{figure}[tbph]
\begin{center}
\includegraphics[bb=18 0 223 143, width=2.5in]{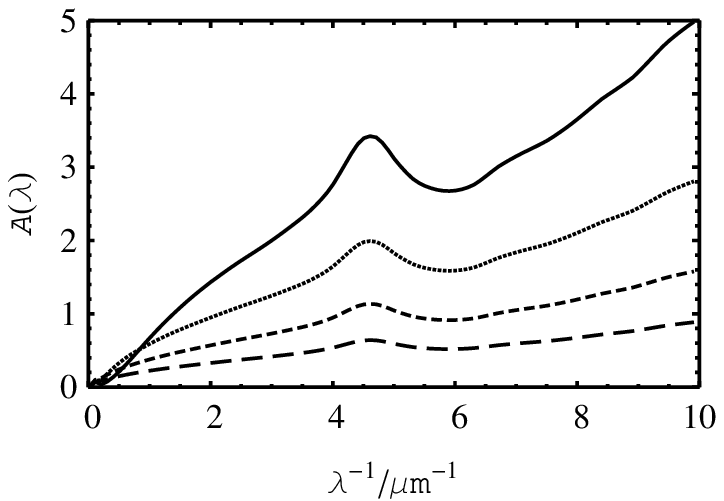}\includegraphics[width=2.5in]{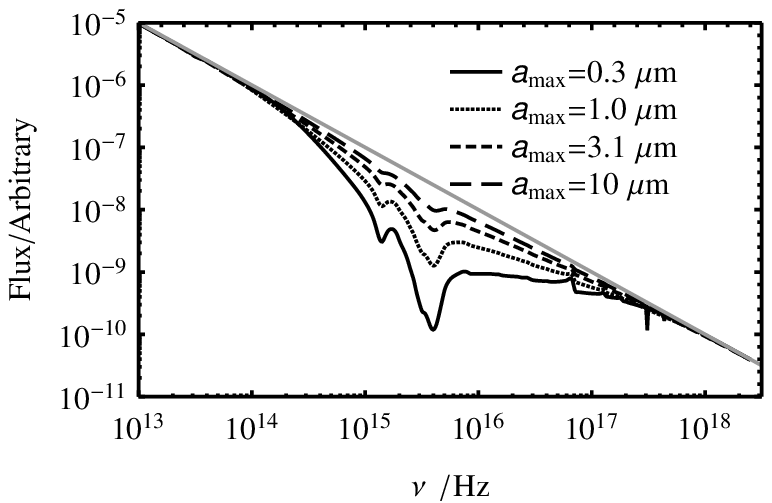}
\end{center}
\caption{Extinction of GRB afterglow by circum stellar
   dust grains with different parameters of size distribution $a_{\rm max}$.
   Left panel is the extinction curve, and right panel is the attenuated afterglow spectrum.
      Gray straight line in the right panel represents the intrinsic spectrum of the afterglow. }
\end{figure}

\section{CONCLUSION} \label{sec:conclusion}

In this work, in order to study the effect of dust extinction on
GRB afterglows, we carry out numerical calculations based on dust
physics and explore the effects of various dust parameters on the
extinction curves. We find that the medium density and the
metallicity determine the absolute amount of extinction, and the
parameters for the size distribution of dust grains $\beta$ and
$a_{\max}$ determine the profile of the extinction curve. When
$\beta$ is larger or $a_{\max}$ is larger, i.e., larger grains are
more excessive, the extinction curve will be flatter with weak
extinction. On the contrary, when $\beta$ is smaller or $a_{\max}$
is smaller, i.e., smaller grains are more excessive, the extinction
curve will be steeper with severe extinction, most likely causing
the optical dark bursts. This may also explain why most bright
afterglows tend to have flatter extinction curves. Therefore, the
massive stellar birth of the GRB and its complex progenitor
environment, should be the major cause of origin of the optical dark
bursts and diverse optical afterglows. Observing and analyzing the
extinction of afterglows would also be important to the studies of
the GRB progenitors and their explosion mechanisms.

\acknowledgments

This work made use of the tabulated
data of interstellar dust provided by B. T. Draine from Princeton
University. We are grateful to S. Aragon from San Francisco State
University for discussion on the MieSolid code and M. A. Caprio from
University of Notre Dame for providing the updated LevelScheme
package and enthusiastic technical support.


\end{document}